\documentclass[fleqn,11pt]{wlpeerj}
\usepackage{hyperref}
\usepackage{url}
\usepackage{geometry}
\usepackage{rotating}
\usepackage{graphics}
\usepackage{multirow}
\usepackage{multirow}
\usepackage{caption}
\usepackage{subcaption}

\title{Hemogram Data as a Tool for Decision-making in COVID-19 Management: Applications to Resource Scarcity Scenarios}

\author[1,2,3,$^\ast$]{Eduardo Avila}
\author[3,4,$^\ast$]{Márcio Dorn}
\author[1,3]{Clarice Sampaio Alho}
\author[3,5,$^\ast$]{Alessandro Kahmann}

\affil[1]{Forensic Genetics Laboratory, School of Health and Life Sciences, Pontifical Catholic University of Rio Grande do Sul, Porto Alegre, RS, Brazil}
\affil[2]{Technical Scientific Section, Federal Police Department in Rio Grande do Sul, Porto Alegre, RS, Brazil}

\affil[3]{National Institute of Science and Technology - Forensic Science, Porto Alegre, RS, Brazil}

\affil[4]{Laboratory of Structural Bioinformatics and Computational Biology,Institute of Informatics, Federal University of Rio Grande do Sul, Porto Alegre, RS, Brazil}

\affil[5]{Institute of Mathematics, Statistics and Physics, Federal University of Rio Grande, Rio Grande, RS, Brazil}

\corrauthor[*]{Eduardo Avila. These authors contributed equally to this work.}{e.avila@edu.pucrs.br; mdorn@inf.ufrgs.br; csalho@pucrs.br; alessandrokahmann@furg.br}

\begin{abstract}
\noindent{\textbf{Background:}} %if any
{COVID-19} pandemics has challenged emergency response systems worldwide, with widespread reports of essential services breakdown and collapse of health care structure. A critical element involves essential workforce management since current protocols recommend release from duty for symptomatic individuals, including essential personnel. Testing capacity is also problematic in several countries, where diagnosis demand outnumbers available local testing capacity.

\noindent{\textbf{Purpose:}} %if any
This work describes a machine learning model derived from hemogram exam data performed in symptomatic patients and how they can be used to predict {qRT-PCR} test results.

\noindent{\textbf{Methods:}} %if any
A \textit{Naïve-Bayes} model for machine learning is proposed for handling different scarcity scenarios, including managing symptomatic essential workforce and absence of diagnostic tests. Hemogram result data was used to predict {qRT-PCR} results in situations where the latter was not performed, or results are not yet available. Adjusts in assumed prior probabilities allow fine-tuning of the model, according to actual prediction context.

\noindent{\textbf{Results:}} 
Proposed models can predict {COVID-19} {qRT-PCR} results in symptomatic individuals with high accuracy, sensitivity and specificity. Data assessment can be performed in an individual or simultaneous basis, according to desired outcome. Based on hemogram data and background scarcity context, resource distribution is significantly optimized when model-based patient selection is observed, compared to random choice. The model can help manage testing deficiency and other critical circumstances.

\noindent{\textbf{Conclusions:}} 
Machine learning models can be derived from widely available, quick, and inexpensive exam data in order to predict {qRT-PCR} results used in {COVID-19} diagnosis. These models can be used to assist strategic decision-making in resource scarcity scenarios, including personnel shortage, lack of medical resources, and testing insufficiency.  
\end{abstract}

\begin{document}

\flushbottom
\maketitle
\thispagestyle{empty}

\section*{Introduction}
\label{S:1}
Since its first detection and description \cite{Huang:2020}, {COVID-19} expansion has brought worldwide concerns to governmental agents, public and private institutions, and health care specialists. Declared as a pandemic, this disease has deeply impacted many aspects of life in affected communities. Relative lack of knowledge about the disease particularities has led to significant efforts devoted to alleviating its effects \cite{Lipsitch:2020}.

Alternatives to mitigate the disease spread include social distancing \cite{Anderson:2020}. Such a course of action has shown some success in limiting contagion rates \cite{tu:2020}. However, isolation policies manifest drawbacks as economic impact, with significant effects on macroeconomic indicators and unemployment rates \cite{nicola:2020}. To address this, governments worldwide have proposed guidelines to manage the essential workforce, considered pivotal for maintaining strategic services and provide an appropriate response to the pandemics expansion \cite{Black:2020}.

Widespread reports of threats to critical national infrastructure have been presented, with significant impact associated with medical attention \cite{Kandel:2020}. Significant pressure is being faced by emergency response workers, with some countries on the brink of collapse of their national health systems \cite{Tannem:2020}. The main concern  associated with {COVID-19} is the lack of extensive testing capacity. Shortage of diagnostic material and other medical supplies pose as a major restraining factor in pandemics control \cite{Ranney:2020}.

The most common {COVID-19} symptoms are similar to other viral infectious diseases, making the prompt clinical diagnostic impractical \cite{Adhikari:2020}. Official guidelines emphasize the use of quantitative real-time {PCR} ({qRT-PCR}) assays for detection of viral {RNA} in diagnosis as the primary reference standard \cite{Tahamtan:2020}. In Brazil, test results are hardly available within at least a week, forcing physicians and health care providers to take strategic decisions regarding patient care without quality information.

Previous reports have described alterations in laboratory findings in {COVID-19} patients. Hematological effects include leukopenia, lymphocytopenia and thrombocytopenia, while biochemical results show variation on alanine and aspartate aminotransferases, creatine kinase and D-dimer levels, among other parameters \cite{Guan:2020,Huang:2020}. Some efforts have been applied to evaluate clinical and epidemiological aspects of this disease using computational methods, such as diagnosis, prognosis, symptoms severity, mortality, and response to different treatments. A useful review of some of these methods is presented by Wynants and collaborators \cite{Wynants:20}. 

The main objective of this article is to provide insights to healthcare decision-makers facing scarcity situations, as a shortage of test capacity or limitations in the essential workforce. A useful method of doing so is using hemogram test results. This clinical exam is widely available, inexpensive, and fast, applying automation to maximize throughput. To do so, we have analyzed hemogram data from Brazilian symptomatic patients with available test results for {COVID-19}. We propose a framework using a \textit{Naïve-Bayes} model for machine learning, where test conditions can be adjusted to respond to actual lack of resources problems. Finally, four distinct scarcity scenarios examples are presented, including handling of the essential workforce and shortage of testing and treatment resources.

\section*{Material and Methods}

\subsection*{Data Collection}
5644 patients admitted to the emergency department of \textit{Hospital Israelita Albert Einstein} ({HIAE} - São Paulo, Brazil) presenting {COVID-19}-like symptoms were tested via {qRT-PCR}. A total number of 599 patients (10.61\%) presented positive results for {COVID-19}. The full dataset contains patients anonymized {ID}, age, {qRT-PCR} results, data on clinical evolution, and a total of 105 clinical tests. Not all data was available for all patients, therefore the number of missing information is significant. All variables were normalized to maintain {anonymity} and remove scale effects. No missing data imputation was performed during model generation to avoid \textit{bias}. Considering the significant ammount of missing data, only 510 patients presented values for all 15 parameters evaluated in hemogram results (comprising the following cell counts or hematological measures: hematocrit, hemoglobin, platelets, mean platelet volume, red blood cells, lymphocytes, leukocytes, basophils, eosinophils, monocytes, neutrophils, mean corpuscular volume ({MCV}), mean corpuscular hemoglobin ({MCH}), mean corpuscular hemoglobin concentration ({MCHC}), and red blood cell distribution width ({RDW}). Data for the above parameters were used in model construction, along with {qRT-PCR} {COVID-19} test results. The full dataset is available in \url{https://www.kaggle.com/einsteindata4u/covid19}.

\subsection*{Machine Learning Analysis - \textit{Naïve Bayes} Classifier}

Machine learning ({ML}) is a field of study in computer science and statistics dedicated to the execution of computational tasks through algorithms that do not require explicit instructions but instead rely on learning patterns from data samples to automate inferences \cite{Mitchell:1997}. These algorithms can infer input-output relationships without explicitly assuming a pre-determined model  \cite{geron:2017,Hastie:2009}. There are two learning paradigms: supervised and unsupervised. Supervised learning is a process in which the predictive models are constructed through a set of observations, each of those associated with a known outcome (\textit{label}). In opposition, in unsupervised learning, one does not have access to the labels, it can be viewed as the task of "spontaneously" finding patterns and structures in the input data. 

Our objective with this study is to predict in advance the results of the {qRT-PCR} test with machine learning models using data from hemogram tests performed on symptomatic patients. The main process can be divided into four steps: (1) \textit{pre-processing of the data} (2)  \textit{selection of an appropriate classification algorithm}, (3)  \textit{model development and validation}, i.e., the process of using the selected characteristics to separate the two groups of subjects (positive for {COVID-19} vs. negative for {COVID-19} in {qRT-PCR} test), and (4) test generated model with additional data. Steps are detailed as follows:

\noindent\textit{Data Pre-processing:} Samples presenting a missing value in any of the 15 evaluated features were removed. A total of 510 patients (73 positives for {COVID-19} and 437 negatives) presented complete data and were considered for the model construction. 

\noindent\textit{Classification Algorithm:} In this work, we use the \textit{Naïve Bayes} ({NB}) classifier, which is a probabilistic machine learning model used for classification tasks. The main reasons for choosing this classifier are due to their low computational cost and clear interpretation. In medicine, the first computer-learn attempts in decision support were based mainly on the Bayes theorem, in order to aggregate data information to physicians' previous knowledge \cite{Martin:1960}. The \textit{Naïve Bayes} ({NB}) method combines the previous probability of an event (also called \textit{\textit{prior} probability}, or simply \textit{\textit{prior}}) with additional evidence (as, for example, a set of clinical data from a patient) to calculate a combined, conditional probability that includes the \textit{\textit{prior}} probability given the extra information. The result is the \textit{posterior probability} of an outcome, or simply \textit{posterior}. This classifier is called "naïve" because it considers that each exam result (variables) is independent of each other. Once this situation is not realistic in medicine, the model should not be interpreted \cite{Schurink:2005}. Besides this drawback, it can outperform more robust alternatives in classification tasks, and once it reflects the uncertainty involved in the diagnosis, Bayesian approaches are more suitable than deterministic techniques  \cite{Gorry:1968,Hastie:2009}.

\noindent\textit{Model Development and Validation:} A classifier is an estimator with a predict method that takes an input array (test) and makes predictions for each sample in it. In supervised learning estimators (our case), this method returns the predicted labels or values computed from the estimated model (positive or negative for {COVID-19}). \textit{Cross-validation} is a model evaluation method that allows one to evaluate an estimator on a given dataset reliably. It consists of iteratively fitting the estimator on a fraction of the data, called training set, and testing it on the left-out unseen data, called test set. Several strategies exist to partition the data. In this work, we used the \textit{Leave-one-out} ({LOO}) cross-validation model, as in Chang et al. \cite{Chang:2003}. The number of data points was split N times (number samples). The method was trained on all the data except for one point, and a prediction was made for that point. The proposed approach was implemented in \textit{Python v.3} (\url{https://www.python.org}) code using \textit{Scikit-Learn v. 0.22.2} \cite{Pedregosa:2011} as a backend. 

\noindent\textit{Model Test:} In order to evaluate the adequacy and generalization power of the proposed model, a set of 92 samples (10 positives for {COVID-19} and 82 negatives) was extracted from the patient database. Those samples were not initially employed in model delineation, considering they present a single missing value among all 15 employed hemogram parameters. Missing data for this training set was imputed using the average value of the missing parameter within the resulting group (positive or negative). The test set was submitted to the previously generated model in order to evaluate classification performance.

\section*{Results}

\subsection*{Descriptive Analysis}

For data description, probability density function ({PDF}) of all 15 hemogram parameters were estimated through the original sample by kernel density estimator. Some hemogram parameters present notable differences between the distributions of positive and negative results, mainly regarding its modal value (distribution peak value) and variance (distribution width). Differences are summarized in Table \ref{tab1}. Regarding basophiles, eosinophils, leukocytes and platelets counts, {qRT-PCR} positive group distribution shows lower modal value and lower variance. On the other hand, monocyte count displays opposite behavior, once lower modal value and variance are observed for the {qRT-PCR} positive group. Lower variance may depict a condition pattern, therefore it is expected that negative cases present higher variance once it may contain a higher variety of conditions (reasons for symptom presence). The remaining nine hemogram parameters did not show a notable difference between negative and positive groups. {PDF} analysis results are presented in Supplementary Material Figure S1.

\begin{table}[htb!]\centering
\caption{Descriptive analysis of hemogram parameters used in present study.}\label{tab1}
\begin{tabular}{lcc}
\hline\hline
\textbf{Parameter} & \textbf{Modal value}        & \textbf{Variance}           \\ \hline\hline
Basophiles         & Reduced in positive cases   & Reduced in positive cases   \\
Eosinophiles       & Reduced in positive cases   & Reduced in positive cases   \\
Leukocytes         & Reduced in positive cases   & Reduced in positive cases   \\
Monocytes          & Augmented in positive cases & Augmented in positive cases \\
Platelets          & Reduced in positive cases   & Reduced in positive cases   \\ \hline\hline
\multicolumn{3}{l}{Parameters not shown displayed no difference between groups}                                  \\ \hline\hline
\end{tabular}
\end{table}

\subsection*{\textit{Naïve Bayes} Model Results} 
\begin{figure}[htb!]
      \centering
          \centering
          \includegraphics[width=\textwidth]{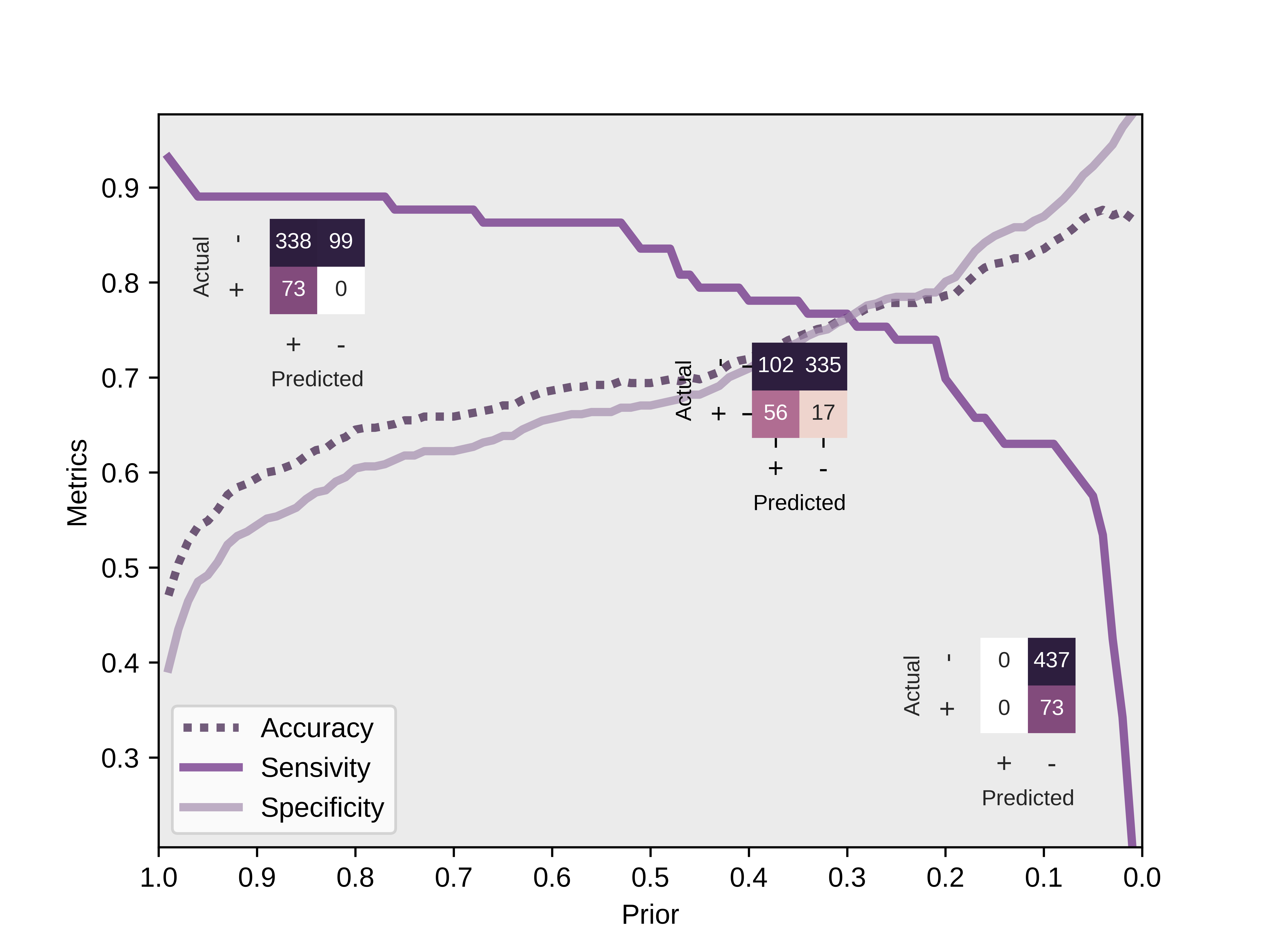}
        \caption{\textbf{Performance metrics of proposed Naive-Bayes model.} \textit{prior} probabilities are presented in reference to positive qRT-PCR prediction. Confusion matrices (left to right) are presented for 0.9999, 0.2933 and 0.0001 \textit{prior} probabilities, respectively. Sensitivity=True Positive Ratio; Specificity=True Negative Ratio}
          \label{fignbr}
\end{figure}

A {NB} classifier based on training set hemogram data was developed. Under the model, the complete range of \textit{prior} probabilities (from 0.0001 to 0.9999 by 0.0001 increments) was scrutinized, and \textit{posterior} probability of each class was computed for different \textit{prior} conditions. A  \textit{posterior} probability value of 0.5 was defined as the classification threshold in one of the positive or negative predicted groups. Resulting model showed a good predictive power of the {qRT-PCR} test result based on hemogram data. Figure \ref{fignbr} shows the accuracy, sensitivity, and specificity curves derived from the model for different \textit{prior} probabilities of each class (positive or negative for {COVID-19}). Reported \textit{prior} probabilities refer to positive {COVID-19} condition.

When setting the \textit{\textit{prior}} probability to the maximum defined value (0.9999), the {NB} classifier correctly diagnosed all {PCR} positive cases. On the other hand, such configuration improperly predicted 77.3\% of negative {PCR} results as positive. Regarding the lower possible \textit{\textit{prior}} probability setting, it does not classify a single observation as positive. This result can be explained by the unbalanced number of observations for each class, tending to over classify samples as the class with more observation, i.e. negative results. Such characteristics can also be noticed in the general accuracy, since the decrease in the \textit{prior} ponce the classifier tends to diagnose all observations as belonging to the dominant class (negative) and consequently raising the total of correctly classified samples. The break-even point is met when \textit{prior} probability is set to 0.2933. Under this condition, all metrics are approximately 76.6\%.

Regarding the model sensitivity, the rate of positive samples correctly classified is over 85\% within 0.999 to 0.5276 range, with small decrease of it when the \textit{prior} probability of positive result is diminished within this range. When {\textit{prior}} is set to under 0.0606, the number of positive predicted samples decrease rapidly, yielding lower sensitivity. As for specificity, it presents linear growth as tested \textit{priors} decrease. Ultimately, the accuracy results profile are similar to specificity, due to the negative patients dominance.   

%(ADICIONAR REFERÊNCIA ÀS FIGURAS)

\begin{figure}[htb!]
      \centering
          \centering
          \includegraphics[width=\textwidth]{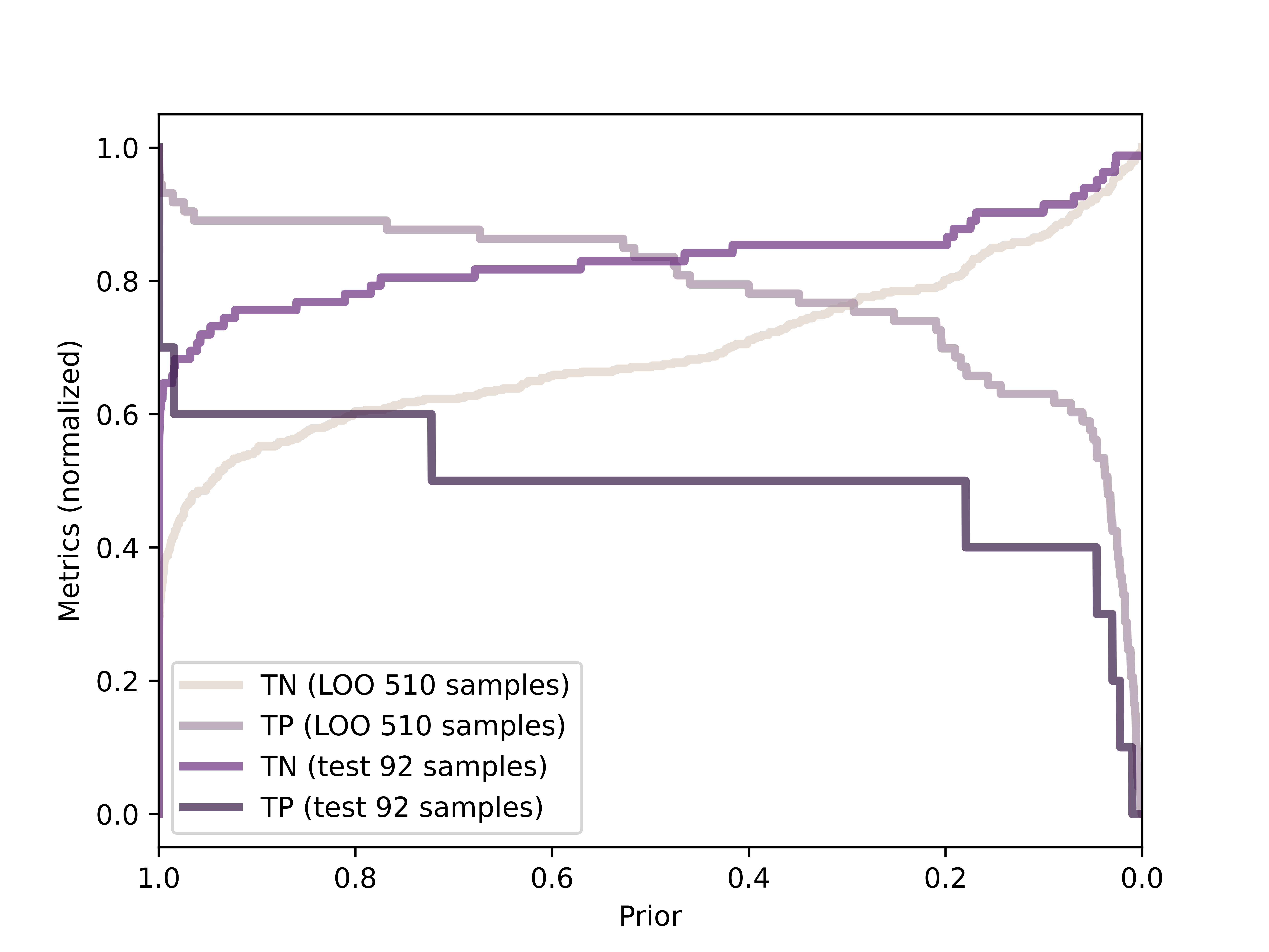}
        \caption{\textbf{Classification performance for training ({{LOO}}) and test datasets.} Results presented for the complete \textit{prior} probability range. Results are presented as the percentage of correctly predicted {qRT-PCR} exams. Informed \textit{prior} probability refers to positive outcome. {TN}: true negative; {TP}: true positive.}
          \label{fig3}
\end{figure}

As mentioned above, \textit{prior} probability choice has a critical relevance in proposed model use. It is clear that, when extreme values of positive probability are applied (close to 0 or 1), specific classes (positive or negative {qRT-PCR} test results predictions) are favoured, increasing its ability of correct detection. As an example, when a value of 0.9999 is set for \textit{\textit{prior}} probability of positive result is set, an increase in misclassification in negative class results is observed. At the same time, it is possible to properly identify samples where hemogram evidence strongly indicates a negative result, according to the model. This is based on the fact that evidence used in the model construction (in present case, hemogram data) must strongly support the reduction of \textit{posterior} probability of disease to values under 0.5, therefore leading to a negative result. This logic can be applied to fine tune the \textit{prior} probability used in the model, in order to improve correct classification of positive or negative groups prediction. Examples of how to use this feature is provided in the “Discussion” section. Test samples (n=92, including 10 {qRT-PCR} positives) were used to test the proposed model. Figure \ref{fig3} presents results obtained from the model application to test dataset.

	\section*{Discussion}

\begin{figure}[htb!]
      \centering

          \centering
          \fbox{\includegraphics[width=0.95\textwidth]{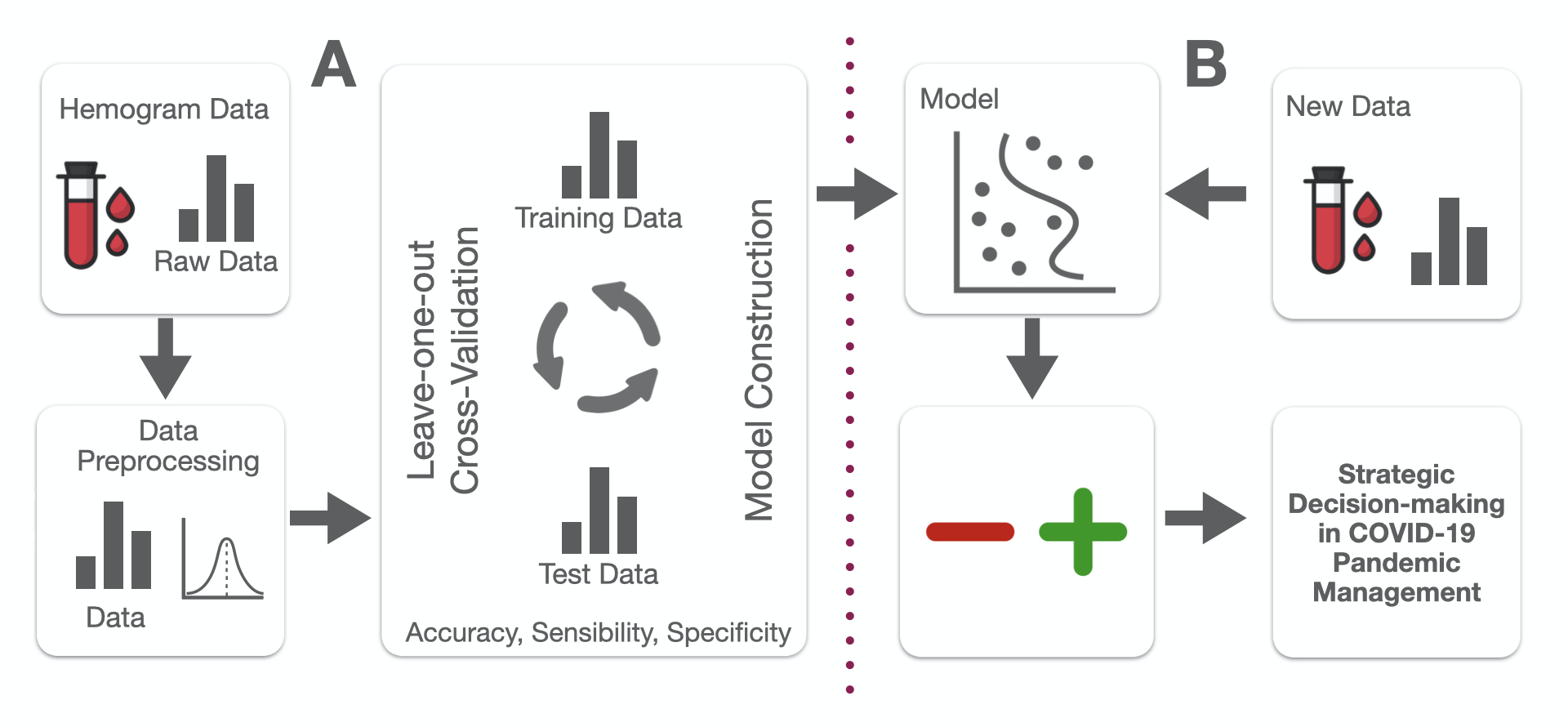}}
          \caption{\textbf{NB Model construction and application diagram.}}
          \label{figmodelo}
\end{figure}

Laboratory findings can provide vital information for pandemics surveillance and management \cite{Lippi:2020}. Hemogram data have been previously proposed as useful parameters in diagnosis and management of viral pandemics \cite{Shimoni:2013}. In the present work, an analysis concerning hemogram data from symptomatic patients suspected of {COVID-19} infection was executed. A machine learning model based on \textit{Naïve Bayes} method is proposed in order to predict actual {qRT-PCR} from such patients. The presented model can be applied to different situations, aiming to assist medical practitioners and management staff in key decisions regarding this pandemic.  Figure \ref{figmodelo} summarizes model construction and application. Predictions are not intended to be used as a diagnostic method since this technique was designed to anticipate {qRT-PCR} results only. As such, it is highly dependable on factors affecting {qRT-PCR} efficiency, and its prediction capacity is dependent on the sensitivity, accuracy, and specificity of the original laboratory exam \cite{Sheturaman:2020}. 

Descriptive analysis of hemogram clinical findings shows differences in blood cell counts and other hematological parameters among {COVID-19} positive and negative patient results. Differences are conspicuous among three measures (leukocytes, monocytes and platelets) and more discrete to additional two (basophiles and eosinophiles). It is possible that differences are also present across the complete data spectrum, even though they are not clearly visualized with {PDF} data. These results are in accordance to previous reports of changes in laboratory findings in {COVID-19} infected patients, where conditions as leukopenia, lymphocytopenia and thrombopenia were reported \cite{Fan:2020}. It is important to highlight that data analysis is not sufficient to characterize clinical hematological alterations in evaluated patients (when compared to demographic hematologic parameters data), once data was normalized for the evaluated sample set only. However, even within this particular quota of population (individuals presenting {COVID-19}-like symptoms), differences were found between individuals presenting negative or positive {qRT-PCR} {COVID} test results. The proposed {NB-ML} model can be helpful in accessing different levels of information from hemogram results, through inferring non-evident patterns and parameter relationships from this data.

Bayesian techniques are based on the choice of a \textit{prior} probability of an event (in present case, positive result for {qRT-PCR} test). The method considers actual evidence (hemogram data) to result in a \textit{posterior} probability of the outcome (prediction of a positive result). By changing the selected \textit{prior} probability, we can derive an uncertainty analysis of the model to understand its distribution. Uncertainty can be then applied to adequately adapt the classifier to a particular ongoing context. This option allows the evaluation of different decision-making scenarios concerning diverse aspects of pandemics management. During a crisis situation, measures should be taken seeking to maximize benefits and achieve a fair resource allocation \cite{Emanuel:2020}. To illustrate the model flexibility and how it can be used to help on this matter, a general framework of application is proposed, followed by a simulation of four scenarios where resource scarcity is assumed.

\subsubsection*{Application Framework}

The proposed {NB} model can be applied in two distinct situations. When clinical data is available for a particular patient, it is highly recommended that medical staff determine the \textit{prior} probability on a case-by-case basis. When no clinical or medical data is available, or when decisions regarding resource management involving multiple symptomatic patients are necessary, the model can be used in multiple individuals simultaneously, aiming to identify those with higher probabilities of presenting positive {qRT-PCR} results.

\subsubsection*{Individual Assessment}

Individual risk management and personal evaluation is essential for {COVID-19} response \cite{Gasmi:2020}. Individuals presenting {COVID-19} symptoms are medically evaluated where no {COVID-19} test is available for appropriate diagnosis confirmation. Medical practitioners can determine a probability of disease based on anamnesis, symptoms, clinical exams, laboratory findings and other available data. This probability of infection, as determined by the physician or medical team, can be considered as the \textit{prior} probability. Using hemogram data as input, and informing the \textit{prior} probability of {COVID-19} based on medical findings, the model will consider hemogram data to inform a \textit{posterior} probability, which can be higher or lower than the original, and based on the hemogram alterations caused by the virus infection. It is important that hemogram data would not be included in original medical assessment and \textit{prior} determination, in order to avoid bias and reduce model overfit.

\subsubsection*{Multiple Patients Evaluation}

It can be used in situations where decisions are necessary for resource management including multiple individuals. Choice of a target group (positive or negative {qRT-PCR} result prediction) should be defined. The model can be applied to multiple individuals simultaneously, with the choice of \textit{prior} probability carefully adjusted to result in a specific number of predicted individuals from the target group, according to the desired outcome. This method increases the correct selection of candidates belonging to the target group, when compared to random selection. When additional clinical data is available, or become available later, patients selected during bulk evaluation should be reassessed individually as proposed in the general framework, in order to reduce misclassifications.

\subsubsection*{Applications to Scarcity Scenarios:}
Examples of proposed model use are presented for some specific scarcity scenarios in Table \ref{tb2}. As can be seen, the model sensitivity can be adjusted by selecting \textit{prior} probability employed, according to desired outcome or interest group. \textit{prior} selection should be carefully decided, based on current context or situation proposed, and must consider the classification group where higher accuracy is intended.

\begin{sidewaystable}\caption{\normalsize Strategies for {NB-ML} model applications and symptomatic patient selection in scarcity conditions. Hemogram test results are available for all symptomatic patients. Scenarios proposed for situations where test results are not available (no testing or waiting {qRT-PCR} test results). Prediction results were appraised in a binary form, with positive or negative classification based on \textit{posterior} probability threshold of 0.5. Results are presented in reference to random patient selection.}\resizebox{\columnwidth}{!}{

\label{tb2}
\begin{tabular}{cccccccc}
\hline\hline
\multirow{4}{*}{\textbf{Condition}}                                                        & \multirow{4}{*}{\textbf{Context example}}                                                                                                 & \multirow{4}{*}{\textbf{Objective}}                                                                                                 & \multirow{4}{*}{\textbf{Strategy}}                                                                                                 & \multirow{4}{*}{\textbf{Action}}                                                                                                                      & \multirow{4}{*}{\textbf{\begin{tabular}[c]{@{}c@{}}Starting / \\ fixed \\ \textit{prior}\end{tabular}}} & \multirow{4}{*}{\textbf{\begin{tabular}[c]{@{}c@{}}Results in training \\ set (positive \\ misclassified \\ among cleared)\end{tabular}}} & \multirow{4}{*}{\textbf{\begin{tabular}[c]{@{}c@{}}Results in test set\\ (positive \\ misclassified \\ among cleared)\end{tabular}}} \\
                                                                                           &                                                                                                                                           &                                                                                                                                     &                                                                                                                                    &                                                                                                                                                       &                                                                                                &                                                                                                                                           &                                                                                                                                      \\
                                                                                           &                                                                                                                                           &                                                                                                                                     &                                                                                                                                    &                                                                                                                                                       &                                                                                                &                                                                                                                                           &                                                                                                                                      \\
                                                                                           &                                                                                                                                           &                                                                                                                                     &                                                                                                                                    &                                                                                                                                                       &                                                                                                &                                                                                                                                           &                                                                                                                                      \\ \hline\hline
\multirow{4}{*}{\begin{tabular}[c]{@{}c@{}}Testing \\ shortage\end{tabular}}               & \multirow{4}{*}{\begin{tabular}[c]{@{}c@{}}Testing capacity \\ is limited to \\ a third of \\ candidates only\end{tabular}}               & \multirow{4}{*}{\begin{tabular}[c]{@{}c@{}}Maximize \\ number \\ of infected \\ patients tested\end{tabular}}                      & \multirow{4}{*}{\begin{tabular}[c]{@{}c@{}}Prioritize \\ TP \\ identification\end{tabular}}                                        & \multirow{4}{*}{\begin{tabular}[c]{@{}c@{}}Fine-tune \textit{prior} \\ until positive \\ reach testing \\ capacity\end{tabular}}                               & \multirow{4}{*}{0.5}                                                                           & \multirow{4}{*}{\begin{tabular}[c]{@{}c@{}}130\% increase \\ in actual infected \\ patients tested\\ (\textit{prior}=0.3482)\end{tabular}}         & \multirow{4}{*}{\begin{tabular}[c]{@{}c@{}}100\% increase in \\ actual infected \\ patients tested \\ (\textit{prior}=0.9607)\end{tabular}}   \\
                                                                                           &                                                                                                                                           &                                                                                                                                     &                                                                                                                                    &                                                                                                                                                       &                                                                                                &                                                                                                                                           &                                                                                                                                      \\
                                                                                           &                                                                                                                                           &                                                                                                                                     &                                                                                                                                    &                                                                                                                                                       &                                                                                                &                                                                                                                                           &                                                                                                                                      \\
                                                                                           &                                                                                                                                           &                                                                                                                                     &                                                                                                                                    &                                                                                                                                                       &                                                                                                &                                                                                                                                           &                                                                                                                                      \\ \hline
\multirow{4}{*}{\begin{tabular}[c]{@{}c@{}}Lack of \\ essential \\ workforce\end{tabular}} & \multirow{4}{*}{\begin{tabular}[c]{@{}c@{}}Professionals with\\ high risk of nocosomial \\ or  work-related \\ transmission\end{tabular}} & \multirow{4}{*}{\begin{tabular}[c]{@{}c@{}}Keep symptomatic, \\ non-infected \\ essential \\ workers in duty\end{tabular}}          & \multirow{4}{*}{\begin{tabular}[c]{@{}c@{}}Search for evident \\ non-infected \\ workers (TN \\ identification)\end{tabular}}      & \multirow{4}{*}{\begin{tabular}[c]{@{}c@{}}All workers are\\  considered as \\ infected, unless \\ model says otherwise\end{tabular}}                 & \multirow{4}{*}{0.9999}                                                                        & \multirow{4}{*}{\begin{tabular}[c]{@{}c@{}}19.4\% of total \\ workforce \\ cleared (0\%)\end{tabular}}                                    & \multirow{4}{*}{\begin{tabular}[c]{@{}c@{}}50\% of total \\ workforce \\ cleared (6.5\%)\end{tabular}}                               \\
                                                                                           &                                                                                                                                           &                                                                                                                                     &                                                                                                                                    &                                                                                                                                                       &                                                                                                &                                                                                                                                           &                                                                                                                                      \\
                                                                                           &                                                                                                                                           &                                                                                                                                     &                                                                                                                                    &                                                                                                                                                       &                                                                                                &                                                                                                                                           &                                                                                                                                      \\
                                                                                           &                                                                                                                                           &                                                                                                                                     &                                                                                                                                    &                                                                                                                                                       &                                                                                                &                                                                                                                                           &                                                                                                                                      \\ \hline
\multirow{4}{*}{\begin{tabular}[c]{@{}c@{}}Lack of \\ essential \\ workforce\end{tabular}} & \multirow{4}{*}{\begin{tabular}[c]{@{}c@{}}Professionals with \\ medium to \\ low risk of \\ transmission\end{tabular}}                   & \multirow{4}{*}{\begin{tabular}[c]{@{}c@{}}Keep symptomatic, \\ non-infected \\ essential\\  workers in duty\end{tabular}}          & \multirow{4}{*}{\begin{tabular}[c]{@{}c@{}}Find ideal balance \\ to simultaneously, \\ maximize both \\ TN and TP\end{tabular}}     & \multirow{4}{*}{\begin{tabular}[c]{@{}c@{}}Use intersection of \\ sensitivity and \\ specificity curves \\ from training set\end{tabular}}            & \multirow{4}{*}{0.2933}                                                                        & \multirow{4}{*}{\begin{tabular}[c]{@{}c@{}}69.0\% of total \\ workforce \\ cleared (5\%)\end{tabular}}                                    & \multirow{4}{*}{\begin{tabular}[c]{@{}c@{}}81.5\% of total \\ workforce \\ cleared (6.6\%)\end{tabular}}                             \\
                                                                                           &                                                                                                                                           &                                                                                                                                     &                                                                                                                                    &                                                                                                                                                       &                                                                                                &                                                                                                                                           &                                                                                                                                      \\
                                                                                           &                                                                                                                                           &                                                                                                                                     &                                                                                                                                    &                                                                                                                                                       &                                                                                                &                                                                                                                                           &                                                                                                                                      \\
                                                                                           &                                                                                                                                           &                                                                                                                                     &                                                                                                                                    &                                                                                                                                                       &                                                                                                &                                                                                                                                           &                                                                                                                                      \\ \hline
\begin{tabular}[c]{@{}c@{}}Limited \\ medical \\ access\end{tabular}                       & \begin{tabular}[c]{@{}c@{}}Medical assistance \\ limited to \\ 20\% of symptomatic \\ individuals only\end{tabular}                       & \begin{tabular}[c]{@{}c@{}}Avoid contagion \\ exposure of non-infected \\ patients in  ER during \\ medical assistance\end{tabular} & \begin{tabular}[c]{@{}c@{}}Eliminate non-infected \\ from candidates for \\ medical assistance \\ (TN identification)\end{tabular} & \begin{tabular}[c]{@{}c@{}}Fine-tune \textit{prior} to select \\ most likely negative results. \\ Select remanining set \\ for medical assistance\end{tabular} & 0.5                                                                                            & \begin{tabular}[c]{@{}c@{}}35.6\% decrease in \\ non-infected \\ patients exposure \\ (\textit{prior}=0.0954)\end{tabular}                         & \begin{tabular}[c]{@{}c@{}}18.8\% decrease in \\ non-infected \\ patients exposure \\ (\textit{prior}=0.4652)\end{tabular}                    \\ \hline\hline
\multicolumn{8}{l}{TP: True Positive; TN: True Negative}                                                                                                                                                                                                                                                                                                                                                                                                                                                                                                                                                                                                                                                                                                                                                                                                                                                                                                                                                                                                      \\ \hline\hline
\end{tabular}}
\end{sidewaystable}

High accuracy in {qRT-PCR} result prediction is achieved based on hemogram information only. Further analysis performed on the original data (not shown) suggest that additional clinical results can improve prediction efficiency. This conclusion is in accordance with previous findings suggesting biochemical and immunological abnormalities, in addition to hematologic alterations, can be caused by COVID-19 disease \cite{Henry:2020}. In this context, the relevance of data employed to generate {ML} models is emphasized. The use of large and comprehensive datasets, containing as much information as possible regarding clinical and laboratory findings, symptoms, disease evolution, and other relevant aspects, is crucial in devising useful and adequate models. The development of nationwide or regional databases based on local data is essential, in order to capture epidemiological idiosyncrasies associated with such populations \cite{Terpos:2020}. Also, natural differences in hemogram results from distinct demographic groups (as seen in reference values according to age, sex, or other physiological factors) can aggregate noise to the model, which can be reduced when large database are employed in model construction, and results can be devised for each ethnographic strata.

Despite having high overall accuracy, performance metrics obtained with proposed model show unequal ability to predict positive or negative results. This situation is caused by a significant imbalance in number of samples belonging to each of this {qRT-PCR} result groups in original data. The use of balanced data in machine learning model design is important to assure high prediction quality \cite{Krawczyk:2016}. The option of maintaining original data in model construction was adopted, since it better represents actual {COVID-19} prevalence among symptomatic patients, and therefore seems to represent a more realistic situation. Additional simulations applying a balanced model (data not shown) using positive group oversampling (to compensate its insufficiency in original data) have devised alternative models with superior prediction power. Alternative balanced model results are presented in Supplementary Material Figure S2.Therefore, additional positive samples will be added to the data and used in future model versions.

As a perspective, collection of hemogram results from asymptomatic patients (in addition to symptomatic individuals) can be used to evaluate the utility of this approach on the detection of asymptomatic infections, in order to provide alternatives in diagnostics, especially in a context of testing deficiency. A web-based application was developed by the authors, in which hemogram data can be introduced for a single individual, along with prior probability of infection, based on data used to generate the present model. The online tool is available at \url{http://sbcb.inf.ufrgs.br/covid}. Future implementation will allow the upload of multiple patients simultaneously, and construction or testing of user data-derived models. This service will allow easy access and practical application of the proposed model.

\bibliographystyle{plain}
\bibliography{main}

\setcounter{figure}{0}

\makeatletter 
\renewcommand{\thefigure}{S\@arabic\c@figure}
\makeatother

\vspace{4cm}
\begin{center}\Large{
    Supplementary Material}
\end{center}

\begin{figure}
     \centering
     \begin{subfigure}[b]{0.32\textwidth}
         \centering
         \includegraphics[width=\textwidth]{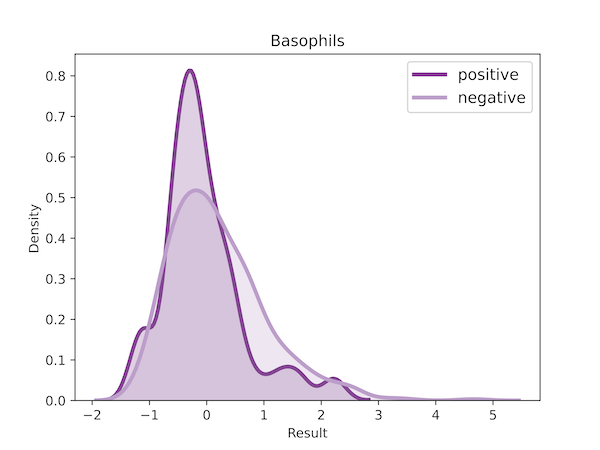}
         \caption{Basophils}
         \label{fig:y equals x}
     \end{subfigure}
     \hfill
     \begin{subfigure}[b]{0.32\textwidth}
         \centering
         \includegraphics[width=\textwidth]{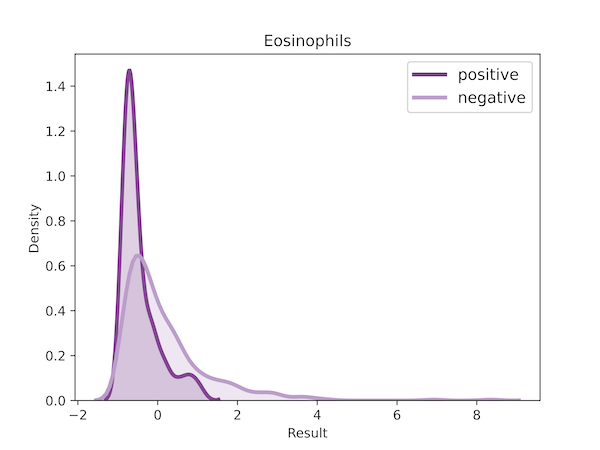}
         \caption{Eosinophils}
         \label{fig:three sin x}
     \end{subfigure}
     \hfill
     \begin{subfigure}[b]{0.32\textwidth}
         \centering
         \includegraphics[width=\textwidth]{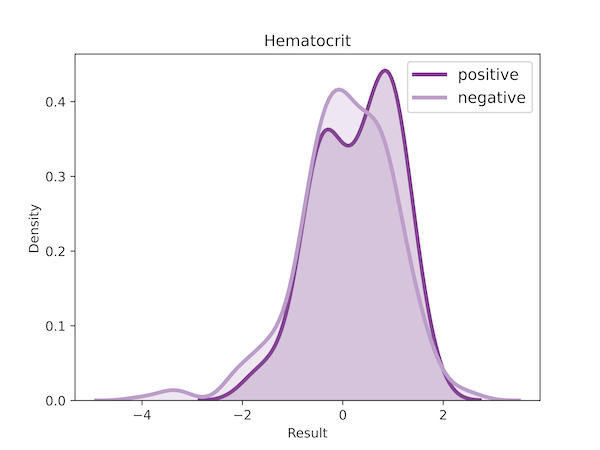}
         \caption{Hematocrit}
         \label{fig:five over x}
     \end{subfigure}

     \begin{subfigure}[b]{0.32\textwidth}
         \centering
         \includegraphics[width=\textwidth]{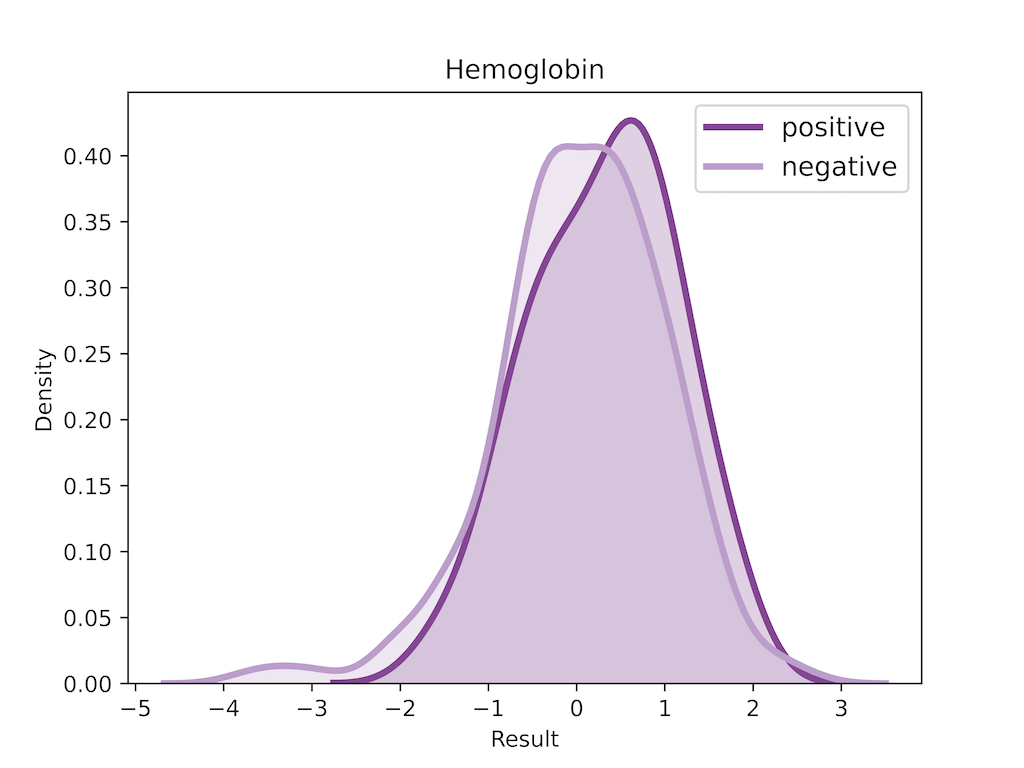}
         \caption{Hemoglobin}
         \label{fig:y equals x}
     \end{subfigure}
     \hfill
     \begin{subfigure}[b]{0.32\textwidth}
         \centering
         \includegraphics[width=\textwidth]{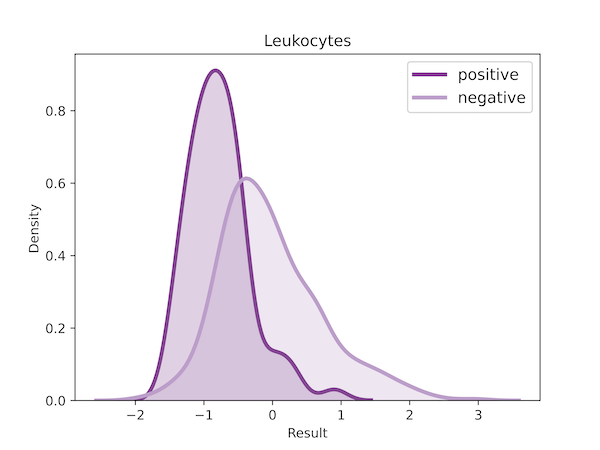}
         \caption{Leukocytes}
         \label{fig:three sin x}
     \end{subfigure}
     \hfill
     \begin{subfigure}[b]{0.32\textwidth}
         \centering
         \includegraphics[width=\textwidth]{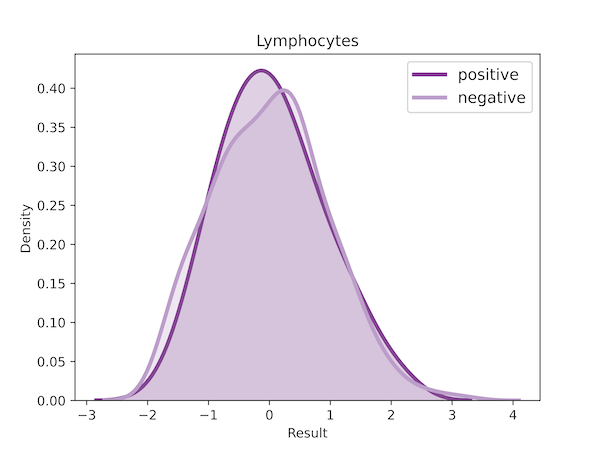}
         \caption{Lymphocytes}
         \label{fig:five over x}
     \end{subfigure}

     \begin{subfigure}[b]{0.32\textwidth}
         \centering
         \includegraphics[width=\textwidth]{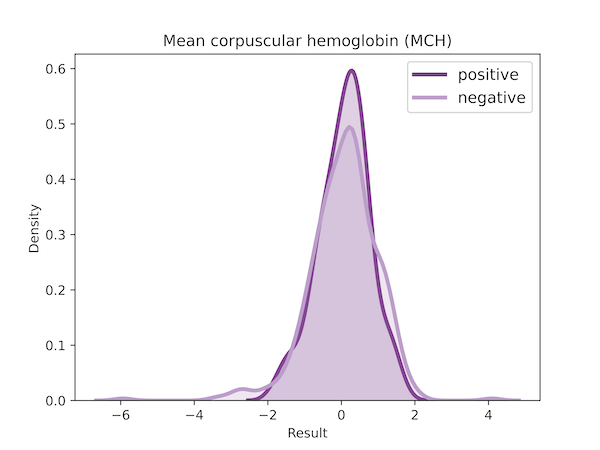}
         \caption{MCH}
         \label{fig:y equals x}
     \end{subfigure}
     \hfill
     \begin{subfigure}[b]{0.32\textwidth}
         \centering
         \includegraphics[width=\textwidth]{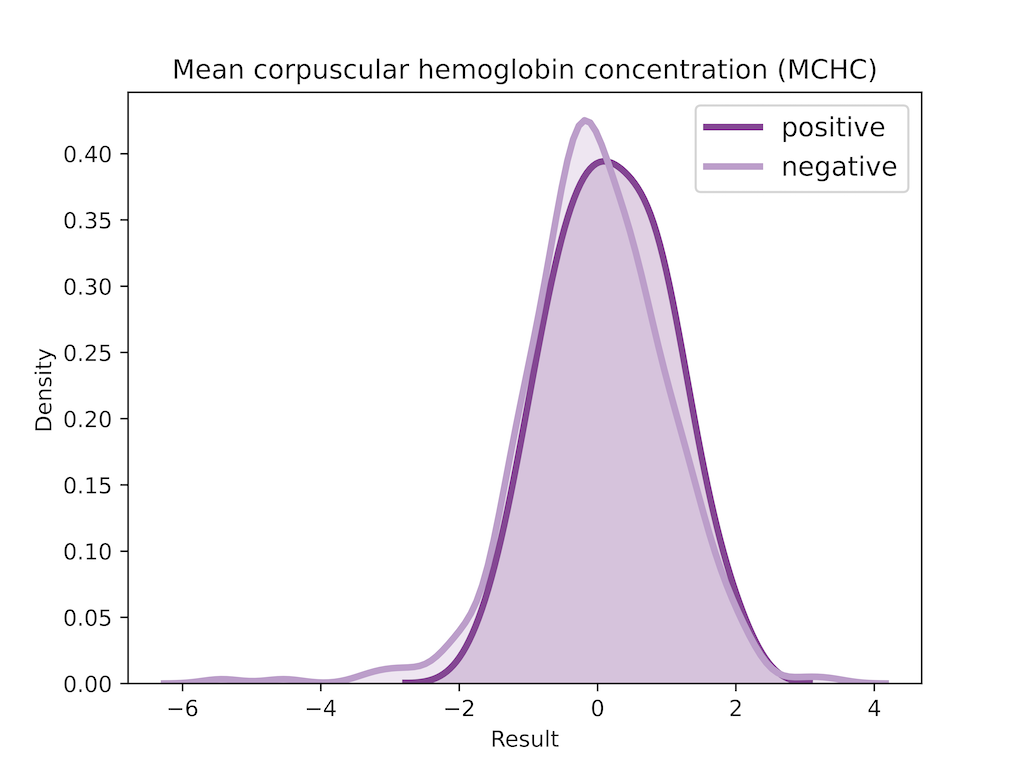}
         \caption{MCHC}
         \label{fig:three sin x}
     \end{subfigure}
     \hfill
     \begin{subfigure}[b]{0.32\textwidth}
         \centering
         \includegraphics[width=\textwidth]{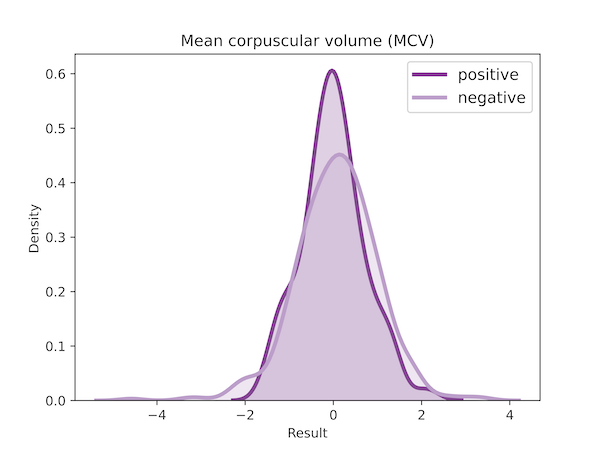}
         \caption{MCV}
         \label{fig:five over x}
     \end{subfigure}

     \begin{subfigure}[b]{0.32\textwidth}
         \centering
         \includegraphics[width=\textwidth]{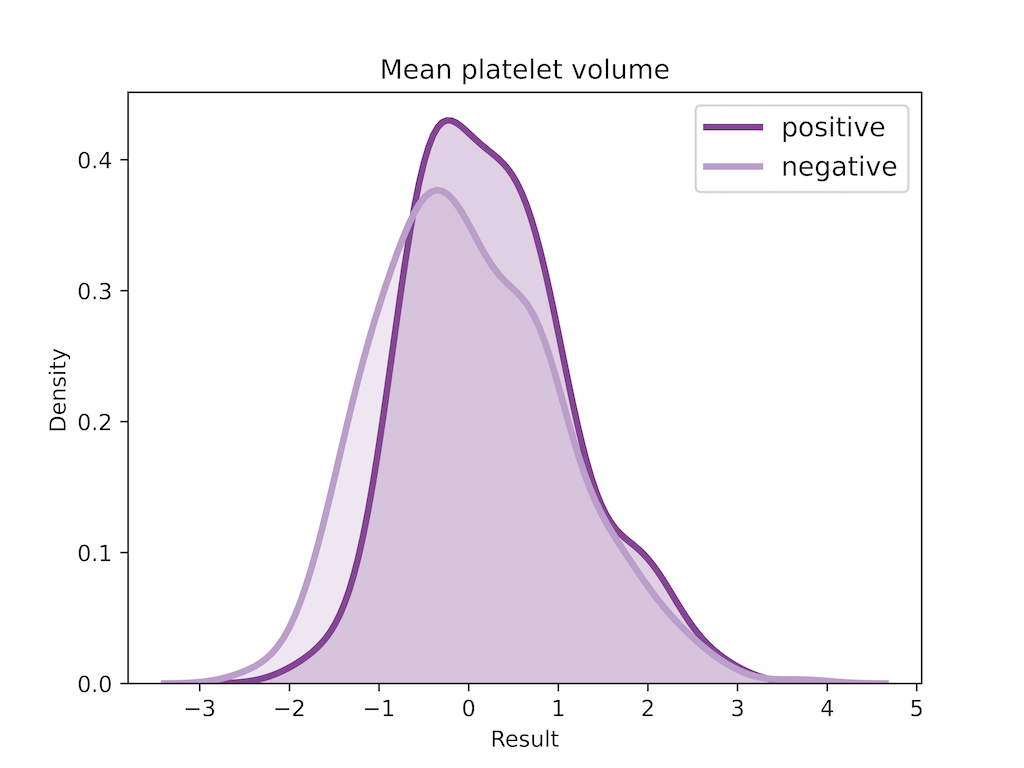}
         \caption{Mean Platelet Volume}
         \label{fig:y equals x}
     \end{subfigure}
     \hfill
     \begin{subfigure}[b]{0.32\textwidth}
         \centering
         \includegraphics[width=\textwidth]{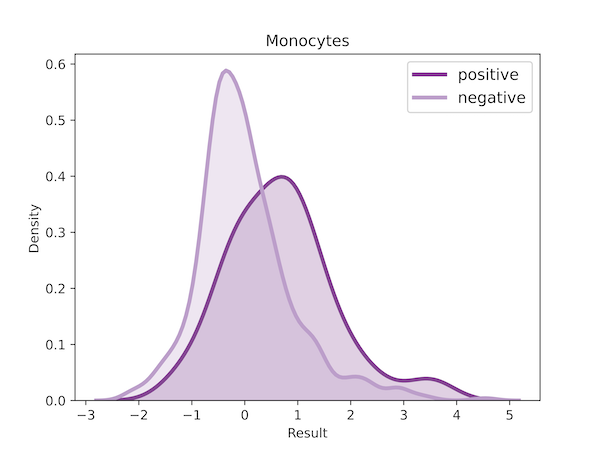}
         \caption{Monocytes}
         \label{fig:three sin x}
     \end{subfigure}
     \hfill
     \begin{subfigure}[b]{0.32\textwidth}
         \centering
         \includegraphics[width=\textwidth]{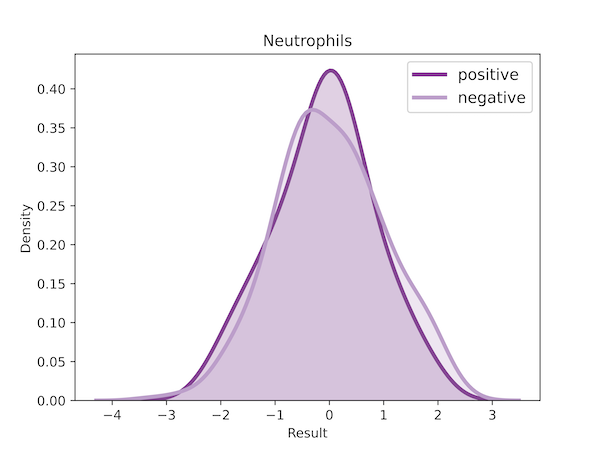}
         \caption{Neutrophils}
         \label{fig:five over x}
     \end{subfigure}

     \begin{subfigure}[b]{0.32\textwidth}
         \centering
         \includegraphics[width=\textwidth]{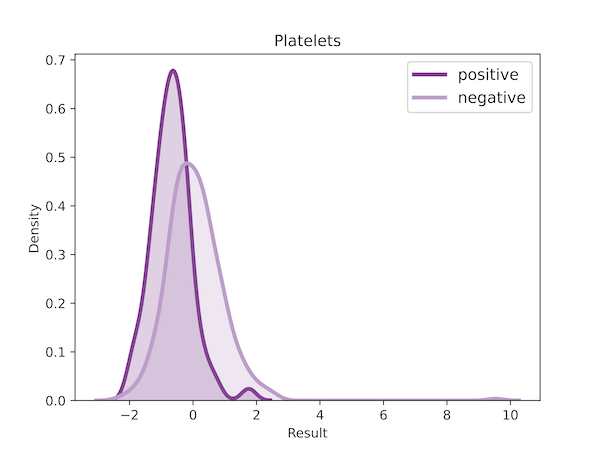}
         \caption{Platelets}
         \label{fig:y equals x}
     \end{subfigure}
     \hfill
     \begin{subfigure}[b]{0.32\textwidth}
         \centering
         \includegraphics[width=\textwidth]{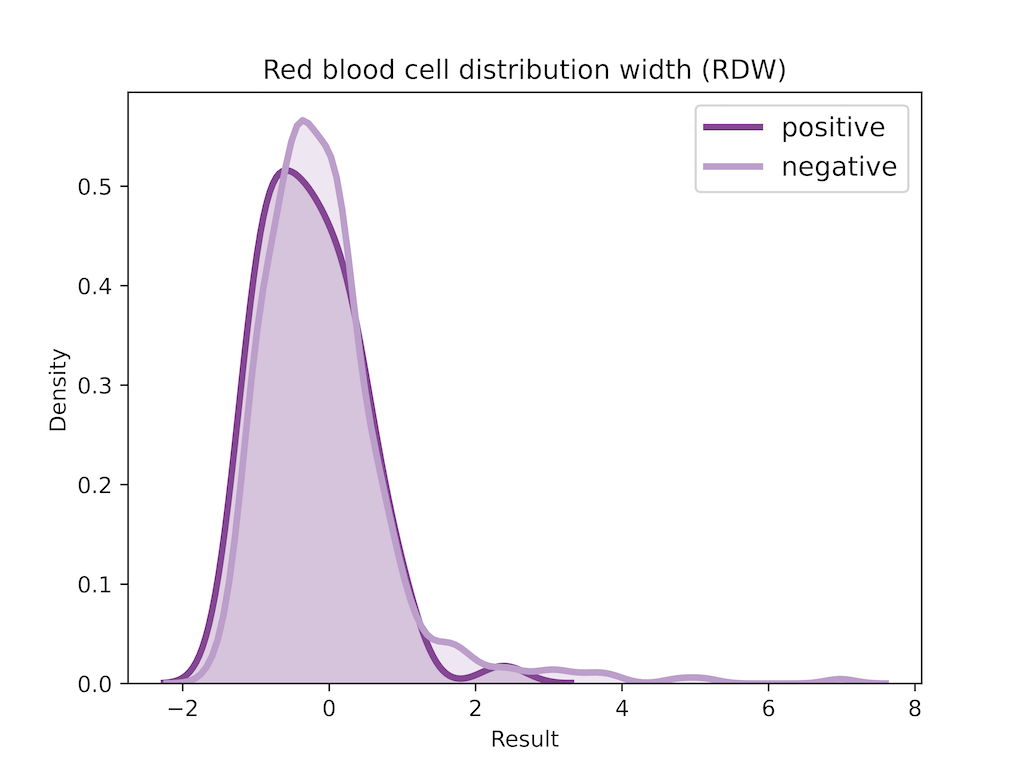}
         \caption{RDW}
         \label{fig:three sin x}
     \end{subfigure}
     \hfill
     \begin{subfigure}[b]{0.32\textwidth}
         \centering
         \includegraphics[width=\textwidth]{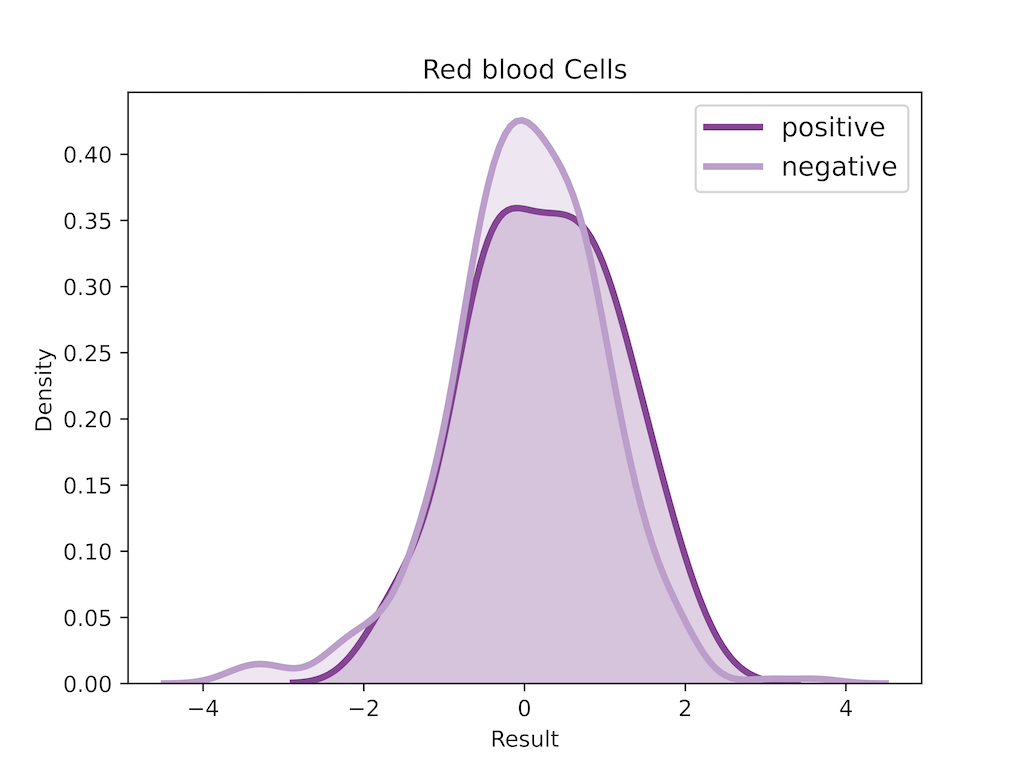}
         \caption{Red Blood Cells}
         \label{fig:five over x}
     \end{subfigure}

        \caption{Probability density function ({PDF}) of all 15 hemogram parameters.}
        \label{fig:three graphs}
\end{figure}

 \begin{figure}[htb!]
      \centering
          \centering
          \includegraphics[width=\textwidth]{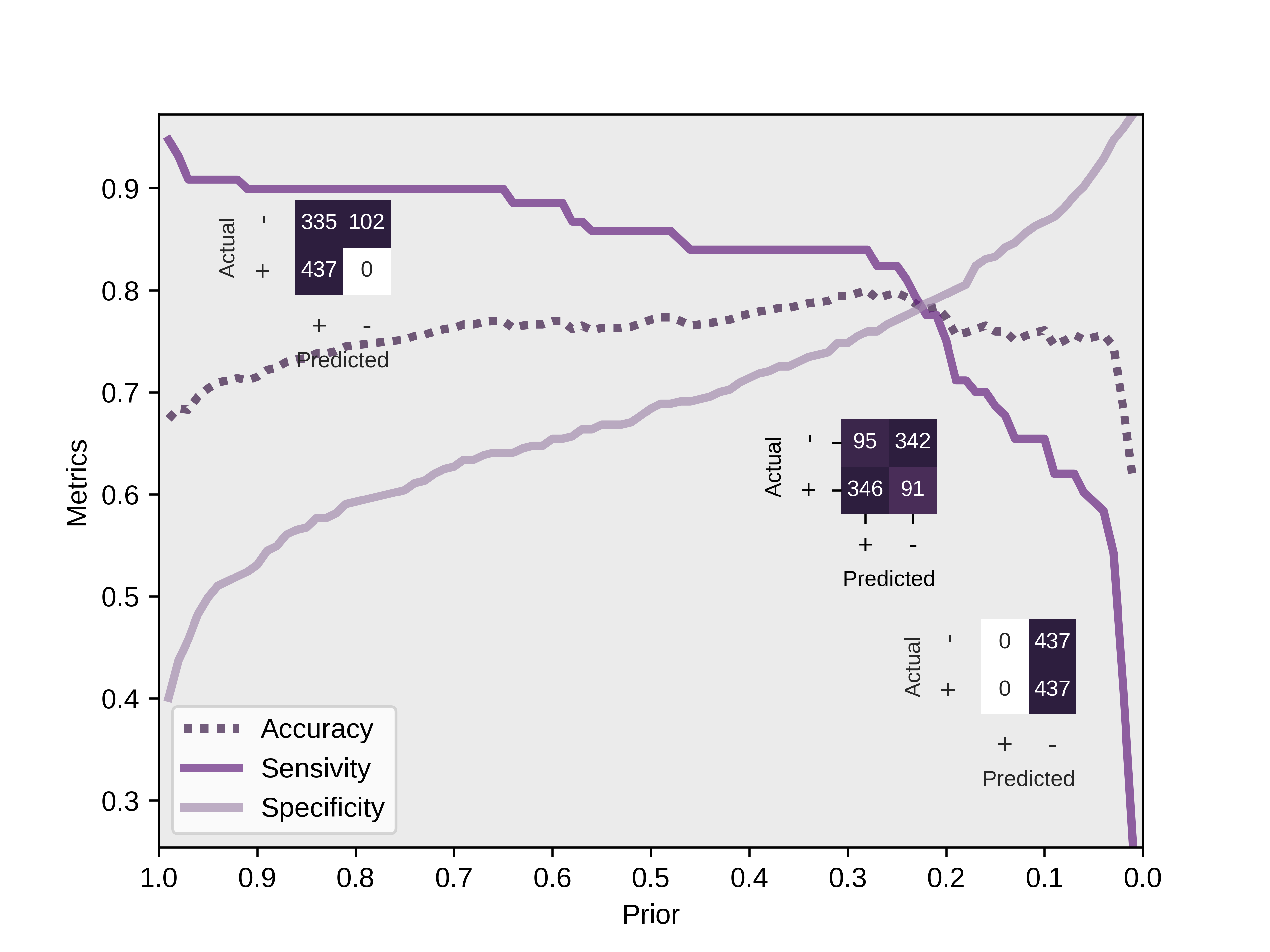}
        \caption{Performance metrics of alternative balanced Naive-Bayes model. In this case, random oversampling of positive results was employed, until sample number in each class is identical. \textit{Prior} probabilities are presented in reference to positive qRT-PCR prediction. Confusion matrices (left to right) are presented for 0.9999, 0.2237 and 0.0001 \textit{prior} probabilities, respectively. Sensitivity=True Positive Ratio; Sensitivity=True Negative Ratio. Random seed was set to 0 for replication purposes.}
          \label{fignbsr}
\end{figure}

\end{document}